# Vector solitons in coupled nonlinear Schrödinger equations with spatial stimulated scattering and inhomogeneous dispersion

E. M. Gromov[1, *], B. A. Malomed[2,3], V. V. Tyutin[1]

[1]National Research University Higher School of Economics, Nizhny Novgorod 603155, Russia

[2]Department of Physical Electronics, School of Electrical Engineering, Faculty of Engineering, Tel Aviv University, Tel Aviv 69978, Israel

[3]Laboratory of Nonlinear-Optical Informatics, ITMO University, St. Petersburg 197101, Russia

**Abstract**

The dynamics of two-component solitons is studied, analytically and numerically, in the framework of a system of coupled extended nonlinear Schrödinger equations, which incorporate the cross-phase modulation, *pseudo-stimulated-Raman-scattering* (pseudo-SRS), cross-pseudo-SRS, and spatially inhomogeneous second-order dispersion (SOD). The system models co-propagation of electromagnetic waves with orthogonal polarizations in plasmas. It is shown that the soliton's wavenumber downshift, caused by pseudo-SRS, may be compensated by an upshift, induced by the inhomogeneous SOD, to produce stable stationary two-component solitons. The corresponding approximate analytical solutions for stable solitons are found. Analytical results are well confirmed by their numerical counterparts. Further, the evolution of inputs composed of spatially even and odd components is investigated by means of systematic simulations, which reveal three different outcomes: formation of a breather which keeps opposite parities of the components; splitting into a pair of separating vector solitons; and spreading of the weak odd component into a small-amplitude pedestal with an embedded dark soliton.

**Highlights**
>Vector solitons in coupled extended nonlinear Schrödinger equations are studied. >We consider media with a pseudo-Raman-scattering and inhomogeneous dispersion. >Analytical and numerical methods are employed. >Soliton solutions are found and investigated. >Evolution of inputs with opposite parities of the two components is explored.

## 1. Introduction

Interest to solitons is bolstered by their ability to travel large distances, keeping the form and transporting encoded data without losses. Soliton solutions are considered in many models of nonlinear propagation of wave fields in dispersive media, which occur in various areas of physics: optical pulses in fibers, electromagnetic waves in plasmas, matter waves in Bose-Einstein condensates, surface waves on deep water, and many others [1-4].

The propagation of broad high-frequency (HF) wave packets is modeled by the second-order nonlinear dispersive wave theory. In isotropic media, the basic equation of the theory is the nonlinear

Schrödinger equation (NLSE) [5,6], which combines the second-order dispersion (SOD) and the self-phase modulation. Its soliton solutions provide for the stable balance between of the dispersive dilatation and nonlinear compression of the wave packets. The consideration of the co-propagation of wave modes with different polarizations is relevant in many settings too, giving rise to coupled NLSEs [7-10], which include cross-phase-modulation (XPM) terms. In this connection, interactions of vector (two-component) solitons in the framework of coupled NLSEs were also studied in detail, see, e.g., [11-13].

The dynamics of narrow HF wave packets is described by the third-order nonlinear dispersive wave theory, which takes into account higher-order terms [1]: nonlinear dispersion (self-steeping) [14], stimulated Raman scattering (SRS) [15-17], and third-order dispersion (TOD). In isotropic media, the basic equation of the theory is the third-order NLSE [17-21]. Soliton solutions were found in the framework of this model, but without the SRS term [22-29]. Such solitons realize the equilibrium of the TOD and nonlinear dispersion. On the other hand, stationary kinks (shock waves) were found as solutions of the third-order NLSE without TOD in works [30,31]. Those solutions are supported by the equilibrium of nonlinear dispersion and SRS. For localized nonlinear wave packets (solitons), SRS leads to the downshift of the soliton spectrum [15-17] and destabilization of the soliton propagation. The use of the balance between the SRS and gain slope for the stabilization of solitons in long telecommunication links was proposed in [32]. The balance between SRS and recoil produced by emission of linear radiation fields from the soliton's core was considered in [33]. In addition, the compensation of the SRS in inhomogeneous media was considered in several settings, *viz.*, those with periodically inhomogeneous SOD [34,35], a shifting zero-dispersion point [36], and dispersion-decreasing fibers [37].

Narrow pulses of HF electromagnetic or Langmuir waves in plasmas, as well as HF surface waves on deep stratified water, suffer induced damping due to scattering on low-frequency (LF) waves in the same medium, which, in turn, are naturally subject to the action of viscosity. These LF modes are ion-sound waves in the plasmas, or internal waves in the stratified fluid. The first model for the damping induced by the interaction with the LF waves was proposed in [38,39]. It gives rise to an extended NLSE with the spatial-domain counterpart of the SRS term, i.e., a *pseudo-SRS* one. The equation was derived from the system of the Zakharov's type [40] for the coupled Langmuir and ion-acoustic waves in plasmas. The pseudo-SRS leads to the self-wavenumber downshift, similar to what is well known in the temporal domain [1,15-17] and, eventually, to destabilization of the solitons. The model equation elaborated in [38,39] also included smooth spatial variation of the SOD due to the medium's inhomogeneity, accounted for by a spatially decreasing SOD coefficient, which leads to an increase of the soliton's wavenumber. The latter property makes it possible to compensate the effect of the pseudo-SRS on the propagation of the soliton by the action of the spatially inhomogeneous SOD. The equilibrium between the pseudo-SRS and decreasing SOD provides stabilization of the soliton's wavenumber spectrum. The soliton dynamics in the model with the nonlinear dispersion and the TOD was considered in [41].

The propagation of bimodal narrow vector wave packets is described by coupled third-order NLSEs, which take into account third-order cross-nonlinear terms [42-45]. In the framework of this system, which does not include SRS terms, two-component vector-soliton solutions were found in [43]. Interactions of vector solitons in the framework of coupled third-order NLSEs were considered in [46].

In the present work we address the dynamics of vector solitons in the framework of coupled extended NLSEs, taking into account pseudo-SRS, cross-pseudo-SRS, XPM and inhomogeneous SOD. Using analytical and numerical methods, the compensation of the soliton's Raman self-wavenumber downshift by the upshift caused by the decreasing SOD is shown. An approximate analytical vector-soliton solution is found in the framework of coupled extended NLSEs, representing the equilibrium of pseudo-SRS and inhomogeneous SOD. By means of direct simulations, we also address the evolution initiated by an input with spatially even and odd components, which reveals different outcomes, depending on the value of the relative amplitude of the two components.

## 2. The model system and integral relations

We consider the propagation of the two-component (vector) HF electromagnetic wave field $\vec{E}(\xi,t) = \left[ U_1(\xi,t)\vec{e}_1 + U_2(\xi,t)\vec{e}_2 \right] \exp(i\omega t - i\kappa\xi)$, where $\vec{e}_{1,2}$ are unit vectors of two orthogonal polarizations, and $U_{1,2}$ are the corresponding amplitudes. The consideration is carried out in the framework of two coupled NLSEs including the pseudo-SRS, cross-pseudo-SRS, XPM and inhomogeneous-SOD terms:

$$2i\left(\frac{\partial U_{1,2}}{\partial t} \mp \delta \frac{\partial U_{1,2}}{\partial \xi}\right) + \frac{\partial}{\partial \xi}\left[q(\xi)\frac{\partial U_{1,2}}{\partial \xi}\right] + 2U_{1,2}\left(|U_{1,2}|^2 + \beta|U_{2,1}|^2\right) + \mu U_{1,2}\frac{\partial\left(|U_{1,2}|^2 + |U_{2,1}|^2\right)}{\partial \xi} = 0, \quad (1)$$

where $q$ is the SOD coefficient, $\delta$ is the group-velocity mismatch between the components, and $\mu$ is the pseudo-SRS strength. These equations are derived from the original Zakharov-type system of equations for two HF electromagnetic waves with orthogonal polarizations and a single LF ion-acoustic field (the LF equation includes the diffusion/viscosity term) in the third-order approximation of the dispersion-wave theory, following the lines of work [38], and taking into account the XPM interaction between the two components of the electromagnetic waves as it was elaborated in detail in optics [47]. In the general case, XPM coefficient $\beta$ in Eq. (1) depends on the exact definition of the two orthogonal polarizations. Below, we consider the case of the XPM of the Manakov's type [48], with $\beta = 1$, which may be realized in the exact form for a special choice of the elliptic polarizations [47]. In fact, the choice of $\beta = 1$ adequately represents the generic situation, as, for instance, one has $\beta = 2/3$ for mutually orthogonal linear polarizations [47], which leads to similar result (for the vector solitons with two components proportional to each other, which is one of types of solutions considered below, all values of $\beta$ are actually mutually equivalent).

Substitution of $U_{1,2} = u_{1,2} \exp\left[\pm i\delta \int d\xi / q(\xi)\right]$ transforms Eqs. (1) into

$$2i\frac{\partial u_{1,2}}{\partial t} + \frac{\partial}{\partial \xi}\left[q(\xi)\frac{\partial u_{1,2}}{\partial \xi}\right] + \frac{\delta^2}{q(\xi)}u_{1,2} + 2u_{1,2}\left(|u_{1,2}|^2 + |u_{2,1}|^2\right) + \mu u_{1,2}\frac{\partial\left(|u_{1,2}|^2 + |u_{2,1}|^2\right)}{\partial \xi} = 0, \quad (2)$$

with an effective potential $\delta^2/q(\xi)$. The definition of the potential implies that $q(\xi)$ does not vanish; it may be interesting too to consider a setting with a zero-dispersion point, at which $q(\xi) = 0$, but in that case it necessary to take into regard the third-order-dispersion term, which will be done elsewhere.

Equations (2) with zero boundary conditions at infinity, $u_{1,2}\big|_{\xi \to \pm\infty} \to 0$, gives rise to the following exact integral relations for a localized wave packet:

$$\frac{dN_{1,2}}{dt} \equiv \frac{d}{dt}\int_{-\infty}^{+\infty}|u_{1,2}|^2 d\xi = 0, \quad (3)$$

$$2\frac{d}{dt}\int_{-\infty}^{+\infty} k_{1,2}|u_{1,2}|^2 d\xi = -\mu\int_{-\infty}^{+\infty}\frac{\partial(|u_{1,2}|^2)}{\partial\xi}\frac{\partial(|u_{1,2}|^2+|u_{2,1}|^2)}{\partial\xi}d\xi - \int_{-\infty}^{+\infty}\frac{dq}{d\xi}\left(\left|\frac{\partial u_{1,2}}{\partial\xi}\right|^2 + \frac{\delta^2}{q^2}|u_{1,2}|^2\right)d\xi$$

$$+2\int_{-\infty}^{+\infty}|u_{1,2}|^2\frac{\partial(|u_{2,1}|^2)}{\partial\xi}d\xi, \qquad (4)$$

$$N_{1,2}\frac{d\bar{\xi}_{1,2}}{dt} \equiv \frac{d}{dt}\int_{-\infty}^{\infty}\xi|u_{1,2}|^2 d\xi = \int_{-\infty}^{+\infty}qk_{1,2}|u_{1,2}|^2 d\xi, \qquad (5)$$

where $u_{1,2}=|u_{1,2}|\exp(i\varphi_{1,2})$, and $k_{1,2}=\partial\varphi_{1,2}/\partial\xi$ are wavenumbers of wave packets $u_{1,2}$.

## 2. Analytical results

### 2.1. Effective evolution equations

To analyze of the wave-packet dynamics, we assume that the scale of the spatial inhomogeneity of SOD is much larger than the packet's width, $D_q \gg \Delta$. Solutions of system (3)-(5) are then looked for in the form of the usual sech ansatz, with two components proportional to each other (solutions of a different type, which represent complexes of even and odd wave forms in the two components, are considered below in a numerical form):

$$u_1(\xi,t) = A(t)\mathrm{sech}\left[\frac{\xi-\bar{\xi}(t)}{\Delta(t)}\right]\exp\left[ik(t)\xi - i\int\Omega(t)dt\right], \quad u_2(\xi,t) = \lambda u_1(\xi,t), \qquad (6)$$

where $\lambda$ is a free real parameter, $\bar{\xi}(t)=\bar{\xi}_{1,2}(t)$ is the coordinate of the center of the moving soliton, $k(t)\equiv k_{1,2}(t)$, $\Delta(t)=(1/A(t))\sqrt{q(\bar{\xi})/(1+\lambda^2)}$, $2\Omega(t)=(1+\lambda^2)A^2(t)+\delta^2/q(\bar{\xi})$, and it is set $A^2(t)\Delta(t)=\mathrm{const}$, which is the usual relation between the amplitude and width of sech-shaped solitons. Substituting ansatz (6) in Eqs. (4) and (5), and taking into account the above conjecture $\Delta \ll D_q$, leads to the following evolution equations:

$$2\frac{dk}{dt} = -\mu\frac{8}{15}\frac{(1+\lambda^2)^2 q_0^2 A_0^4}{q^3(\bar{\xi})} - \frac{(1+\lambda^2)q_0 A_0^2 q'(\bar{\xi})}{3q^2(\bar{\xi})} - \frac{\delta^2 q'(\bar{\xi})}{q^2(\bar{\xi})} - q'(\bar{\xi})k^2, \quad \frac{d\bar{\xi}}{dt} = kq(\bar{\xi}), \qquad (7)$$

where initial values are $q_0 \equiv q(\bar{\xi}(t=0))$, $A_0 \equiv A(t=0)$, which obey the aforementioned relation, $A^2(t)q(\bar{\xi}(t)) = A^2(t=0)q(\bar{\xi}(t=0)) \equiv A_0^2 q_0$, and $q'(\bar{\xi}) \equiv dq/d\xi|_{\xi=\bar{\xi}}$ is the derivative (*slope*) of the SOD coefficient at the soliton's center. Equations (7) give rise to an obvious equilibrium state (alias a fixed point, FP):

$$8\mu(1+\lambda^2)^2 q_0^2 A_0^4 = -5q'(\bar{\xi}_*)q(\bar{\xi}_*)\left[(1+\lambda^2)q_0 A_0^2 + 3\delta^2\right], \quad k_* = 0, \qquad (8)$$

where $\bar{\xi}_*$ is the equilibrium position of the soliton. In the particular case of $\lambda=\delta=0$, relation (8) reduces to its counterpart for the single NLSE derived in [38,39]. For $\mu = \mu_* \equiv -(5/8)q'(\bar{\xi}_0)\left[(1+\lambda^2)q_0 A_0^2 + 3\delta^2\right]/\left[(1+\lambda^2)^2 q_0 A_0^4\right]$ the equilibrium position of the soliton coincides with its initial position, $\bar{\xi}_* = \bar{\xi}_0 \equiv \bar{\xi}(t=0)$.

To analyze the evolution near the FP, we assume a constant value of the SOD slope around the FP, $q' = \text{const}$, and rescale the variables by defining $\tau \equiv -tq'\sqrt{A_0^2 q_0(1+\lambda^2)+3\delta^2}/(\sqrt{3}q_0)$,

$$n(\tau) \equiv q(\bar{\xi}(\tau))/q_0, \quad y(\tau) \equiv k(\tau)\sqrt{3}q_0/\sqrt{A_0^2 q_0(1+\lambda^2)+3\delta^2}, \quad (9)$$

thus deriving a simple mechanical system from Eqs. (7):

$$2\frac{dy}{d\tau} = -\frac{\sigma}{n^3} + \frac{1}{n^2} + y^2, \quad \frac{dn}{d\tau} = -ny, \quad (10)$$

where $\sigma \equiv \mu/\mu_*$. Obviously, Eqs. (10) conserve the corresponding Hamiltonian,

$$2y^2 n^3 - (2y_0^2 + \sigma - 2)n^2 - 2n + \sigma = 0, \quad (11)$$

where $y_0 = y(\tau = 0)$. Dynamical invariant (11) is drawn in the plane of $(y,n)$ in Fig. 1(a), for $y_0 = 0$ and different values of $\sigma$. Trajectories in the plot are closed for $0 < \sigma < 2$, and open otherwise.

Further, at $y_0^2 > 0$ straightforward analysis of Eq. (10) demonstrates that the closed trajectories, which are shown in Fig. 1(a) for $y_0^2 = 0$, stretch in both positive and negative vertical directions (along the axis of $n$). In the same case, the critical value of the pseudo-SRS coefficient, which leads to the destruction of the soliton, decreases to $\sigma_{cr} = 2(1 - y_0^2)$, the destruction being signaled by the disappearance of closed trajectories. Thus, the solitons do not exist at $y_0^2 > 1$; in other words, they exist with the wavenumber smaller than a critical value, $k^2 < A_0^2(1+\lambda^2)/(3q_0) + \delta^2/q_0^2$. Contours of dynamical invariant (11) are plotted in the plane of $(y,n)$ in Fig. 1(b), for $0 < y_0^2 < 1$ and several values of $\sigma$.

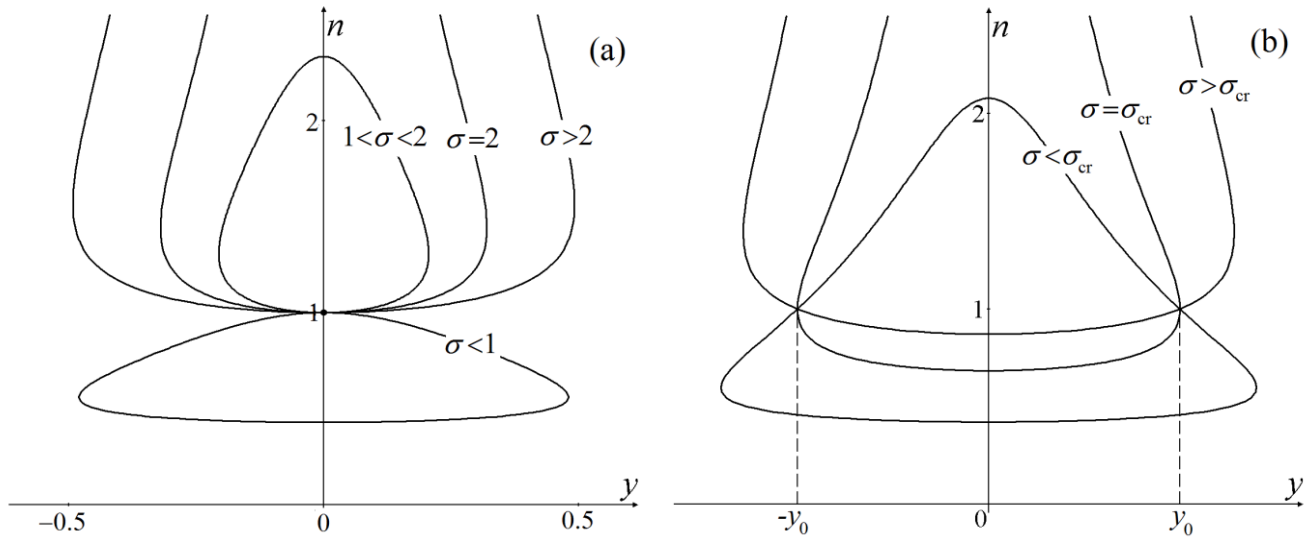

Fig. 1. Contour plots of dynamical invariant (11) in the plane $(y,n)$ of the soliton's rescaled dispersion and wavenumber [see Eqs. (9)], for $y_0 = 0$ (a) and $0 < y_0^2 < 1$ (b), and different values of constant $\sigma$.

## 2.2. The soliton solution

Here we address steady-state solutions of Eqs. (2) for a linear profile of the inhomogeneous SOD, viz., $q(\xi) = q_0 + q'\xi$, in the form of $U_2(\xi,t) = \lambda U_1(\xi,t) \equiv \lambda \psi(\xi)\exp(i\Omega t)$:

$$-2\Omega\psi + \frac{\delta^2}{q_0 + q'\xi}\psi + (q_0 + q'\xi)\frac{d^2\psi}{d\xi^2} + q'\frac{d\psi}{d\xi} + 2(1+\lambda^2)\psi^3 + \mu(1+\lambda^2)\psi\frac{d(\psi^2)}{d\xi} = 0. \quad (12)$$

Similar to what was adopted above, we again assume that the wave-packet's width is much smaller than the scale of the SOD's spatial inhomogeneity, $\Delta \ll 1/|q'|$. Introducing the corresponding small parameter, $\varepsilon \sim \Delta \cdot q' \sim \mu \ll q_0$, a solution to Eq. (12) can be looked for in the standard perturbative form, $\psi = \Phi + \phi$, where $\phi \sim \varepsilon$ is a small correction to $\Phi \sim \varepsilon^0$. Separating terms of orders $\varepsilon^0$ and $\varepsilon^1$, we obtain

$$q_0\frac{d^2\Phi}{d\xi^2} + 2\Phi^3(1+\lambda^2) - \left(2\Omega - \frac{\delta^2}{q_0}\right)\Phi = 0, \quad (13)$$

$$q_0\frac{d^2\phi}{d\xi^2} + \left[6(1+\lambda^2)\Phi^2 - 2\Omega + \frac{\delta^2}{q_0}\right]\phi = q'\frac{\delta^2}{q_0^2}\Phi\xi - q'\frac{d^2\Phi}{d\xi^2}\xi - q'\frac{d\Phi}{d\xi} - \frac{2}{3}\mu(1+\lambda^2)\frac{d(\Phi^3)}{d\xi}. \quad (14)$$

Equation (13) has the standard sech-soliton solution, $\Phi = A\,\text{sech}(\xi/\Delta)$, where $\Delta = \sqrt{q_0/(1+\lambda^2)}/A$, and $2\Omega = (1+\lambda^2)A^2 + \delta^2/q_0$. Then, in terms of rescaled variables, $\eta \equiv \xi/\Delta$ and $\phi \equiv q'\Psi/\sqrt{q_0(1+\lambda^2)}$, Eq. (14) takes the form of

$$\frac{d^2\Psi}{d\eta^2} + \left(\frac{6}{\cosh^2\eta} - 1\right)\Psi = \left[\frac{\delta^2}{q_0(1+\lambda^2)A_0^2} - 1\right]\frac{\eta}{\cosh\eta} + \frac{2\eta}{\cosh^3\eta} + \frac{\sinh\eta}{\cosh^2\eta} + \frac{2\mu(1+\lambda^2)A_0^2}{q'}\frac{\sinh\eta}{\cosh^4\eta}. \quad (15)$$

An essential result is that, at

$$\mu = \mu_* \equiv -(5/8)q'(1+3\text{H})/(1+\lambda^2)A_0^2, \quad (16)$$

where $\text{H} \equiv \delta^2/\left[q_0(1+\lambda^2)A_0^2\right]$, Eq. (15) has an *exact* localized solution for the correction to the standard sech soliton,

$$\Psi(\eta) = (1/4)(\text{sech}\,\eta)\left[2\text{H}\eta + (1-\text{H})\eta^2\tanh\eta - (1+3\text{H})(\tanh\eta)\ln(\cosh\eta)\right]. \quad (17)$$

In the particular case of $\text{H} = 0$, which corresponds to $\delta = 0$, i.e., in the absence of the group-velocity mismatch between the polarization components, solution (17) carries over into one obtained earlier [38,39].

## 3. Numerical results

To check the above analytical results, we here aim to simulate the evolution of initial wave packet $u_{1,2}(\xi,0) = (1/\sqrt{2})\text{sech}\,\xi$ in the framework of Eqs. (2) with a typical linear profile of the inhomogeneous SOD, $q = 1 - \xi/20$, $\delta = 1$, $\lambda = 1$, and different values of strength $\mu$ of the pseudo-SRS effect. The respective point (16) of the equilibrium between the pseudo-SRS and inhomogeneous SOD is $\mu_* = 1/8$. In the simulations performed with $\mu = 1/8$, at times $t > 10$ the pulse evolves into a stationary localized profile with zero wavenumber. Figure 2 shows the deviation of the absolute value of the numerically found stationary profile from the sech-soliton input, i.e., $\phi_{num}(\xi) = |u_{1,2}(\xi)| - (1/\sqrt{2})\text{sech}\,\xi$ (the solid curve in the figure). The deviation is very close to the respective analytically predicted correction, given by Eq. (17):

$$\phi = -(\sqrt{2}/80)(\text{sech}\,\xi)\left[\xi - 2\tanh\xi \ln(\cosh\xi)\right], \tag{18}$$

as shown by the dashed curve in Fig. 2.

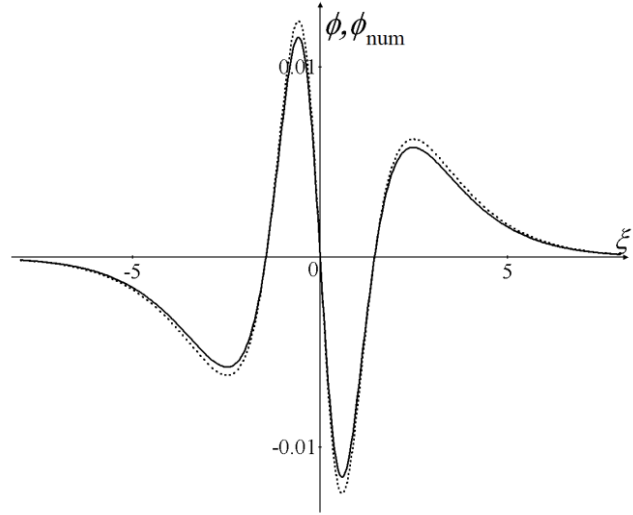

Fig.2. Numerical results: deviation of the absolute value of the numerically found stationary pulse from the standard soliton shape, $\phi_{num}(\xi) = |u_{1,2}(\xi)| - (1/\sqrt{2})\text{sech}\,\xi$ (the solid curve). The analytical correction $\phi$ to the absolute value of the standard soliton solution, given by Eq. (18), is shown by the dashed curve.

Taking values of the pseudo-SRS strength $\mu$ different from the equilibrium point $\mu_* = 1/8$ leads to variation of the soliton's wavenumber and amplitude. In particular, Fig. 3 shows the simulated spatiotemporal evolution of $|u_{1,2}(\xi,t)|$ for $\mu = (4/3)\mu_* \equiv 1/6$. In this case, the soliton performs spatial oscillations without any visible radiation loss, i.e., the soliton is dynamically stable in the case, in the oscillatory state.

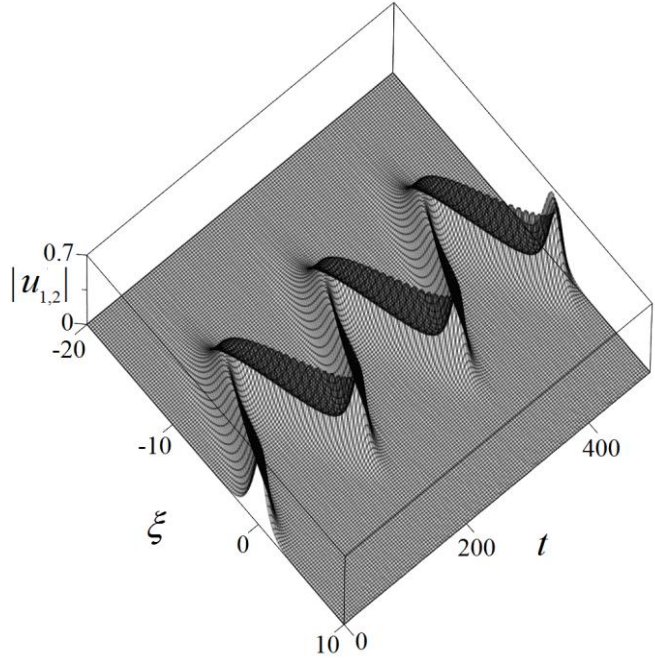

Fig.3. Results of the simulations of the evolution of the sech-shaped pulse for $\mu = (4/3)\mu_* \equiv 1/6$.

The above considerations were focused on two-component solitons with mutually similar shapes of the two components. Another issue of straightforward interest is to consider the evolution of inputs with opposite parities of the components. For this purpose, we carried out the simulations initiated by the input with an even profile in one component, and an odd one in the other:

$$u_1(\xi,0) = \mathrm{sech}\,\xi, \quad u_2(\xi,0) = A[\mathrm{sech}(\xi+1) - \mathrm{sech}(\xi-1)], \tag{19}$$

in the framework of Eqs. (2) with $q = 1 - x/20$, $\delta = 0$, and different values of $A$ and $\mu$. Figures 4-6 display the resulting spatiotemporal evolution of $|u_1(\xi,t)|$ (a) and $|u_2(\xi,t)|$ (b). For the relative amplitude of the odd component $A = 0.8$ (with $\mu = 1/10$), initial pulse (19) transforms into an *essentially novel* dynamical mode, in the form of a breather which keeps the *opposite parities* in its components (Fig.4). On the other hand, for $A = 1$ (with $\mu = 1/35$) initial pulse (19) splits into two separating vector solitons of the usual type, with identical parities in the two components (Fig.5), which is possible as the odd component in Eq. (19), $u_2(\xi,0)$, is built as a set of two pulses with opposite signs. Lastly, for $A = 0.5$ (with $\mu = 1/30$) the weaker component $u_2$ tends to spread out into a small-amplitude pedestal, into which a dark soliton is embedded (Fig. 6(b)), while the even component $u_1$ shows no essential evolution (Fig. 6(a)). In the latter case, the $u_2$ component keeps the spatially odd structure, as dark solitons are odd kink-like solutions.

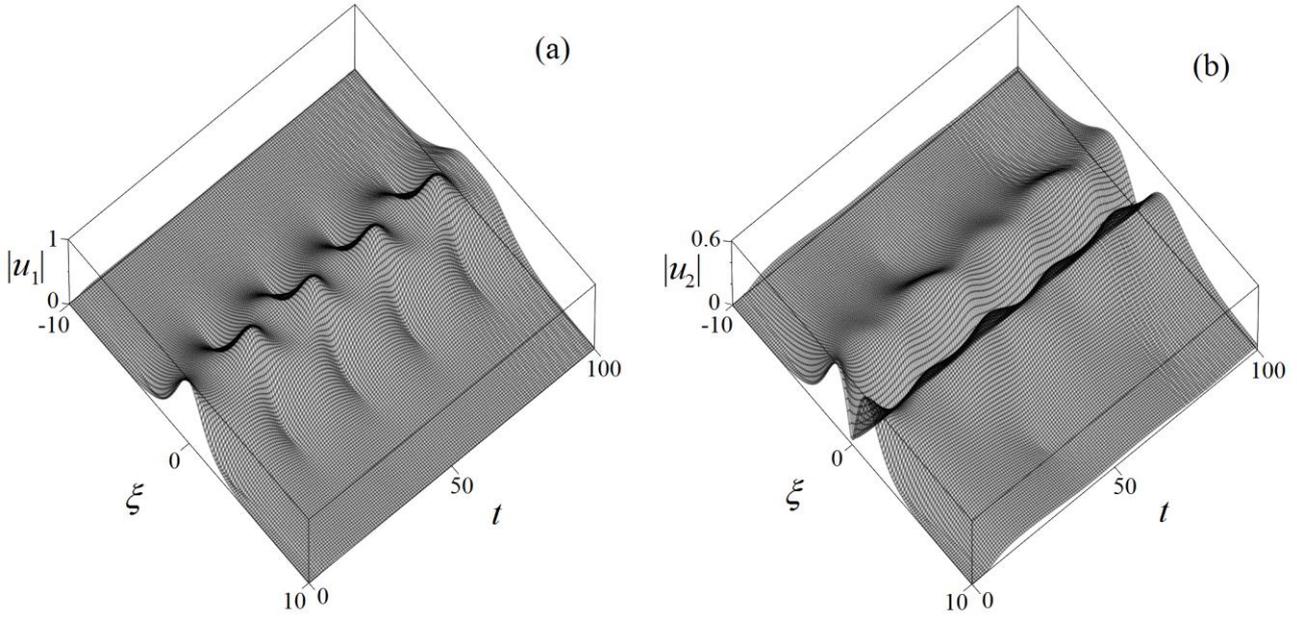

Fig.4. The result of simulations of the evolution of the initial pulse (19) with opposite parities of the components, for $A = 0.8$ and $\mu = 1/10$: formation of a breather with coupled even and odd components.

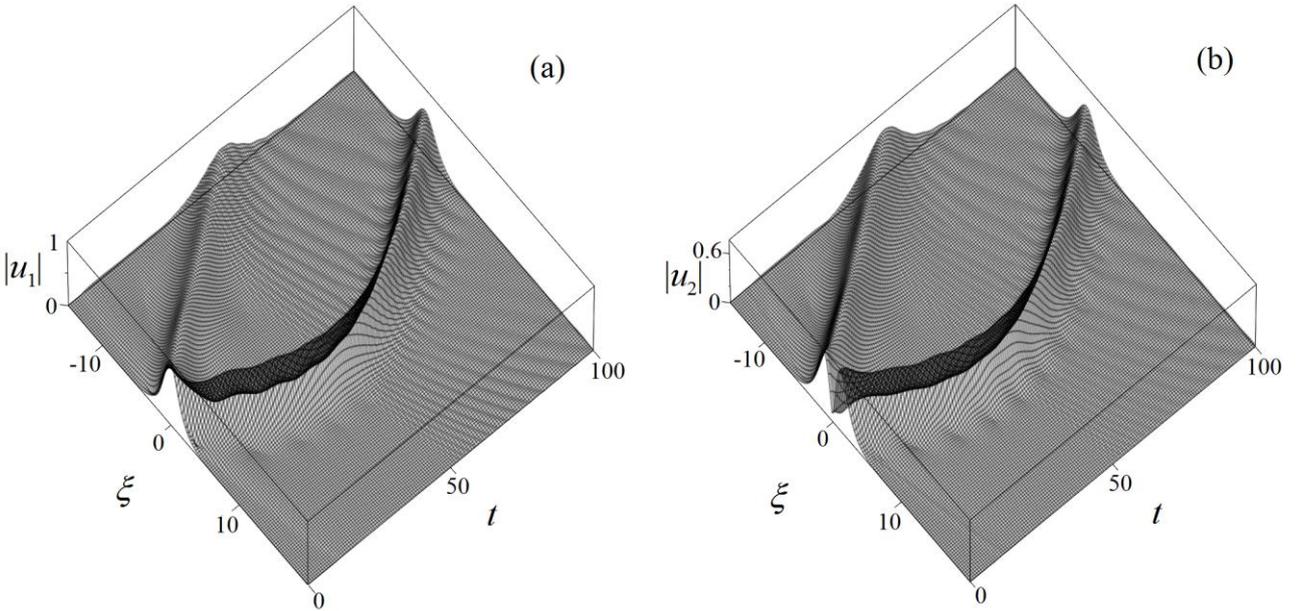

Fig.5. The result of simulations of the evolution of the initial pulse (19) for $A = 1$ and $\mu = 1/25$: splitting into two vector solitons of the usual type.

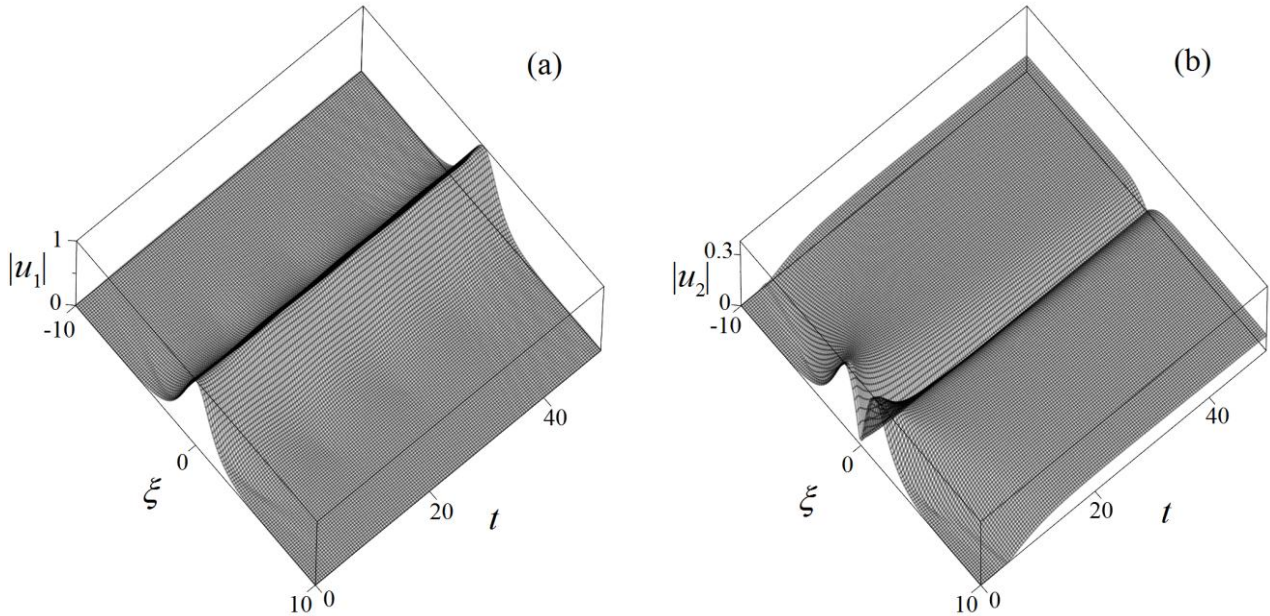

Fig.6. The result of simulations of the evolution of the initial pulse (19) for $A=0.5$ and $\mu=1/30$: the transformation of the weak odd component into a small-amplitude dark soliton.

## 4. Conclusion

We have considered the dynamics of vector (two-component) solitons in the framework of coupled extended NLSEs (nonlinear Schrödinger equations), which take into account essential physical effects, such as the single-component pseudo-SRS (stimulated Raman scattering), cross-pseudo-SRS which couples the components, XPM (cross-phase modulation), and spatially inhomogeneous SOD (second-order dispersion). The compensation of the pseudo-SRS-induced self-wavenumber downshift for the soliton by the downshift induced by the inhomogeneous SOD give rise to stable stationary solitons. A stable analytical soliton solution was obtained, which is very accurately corroborated by numerical results. The critical value of the wave-packet's wavenumber necessary for the existence of the soliton was found. Furthermore, simulations initiated by the input with opposite parities of the two components reveal new outcomes, *viz.*, the formation of breathing bound states of the spatially even and odd components, splitting into a pair of separating vector solitons, and spreading of the weak odd component into a small-amplitude dark soliton.

It may be relevant to include additional effects into the consideration, such as nonlinear dispersion and third-order dispersion. Motion of stable solitons and collisions between them is an interesting possibility too. Extension of the analysis in these directions will be presented elsewhere.

## Acknowledgements

This work was supported by the Russian Foundation for Basic Research Projects, through grant No. 15-02-01919a. The work of B.A.M. is supported, in part, by grant No. 2015616 from the joint program in physics between NSF (US) and Binational (US-Israel) Science Foundation.